\documentclass[epj]{svjour}

\usepackage{graphicx}
\usepackage{amsmath}
\usepackage{amsfonts}
\usepackage{cancel}
\usepackage{lmodern,dsfont}
\usepackage{ulem}
\usepackage[usenames]{color}

\newcommand{\be}{\begin{equation}}
\newcommand{\ee}{\end{equation}}
\newcommand{\ba}{\begin{eqnarray}}
\newcommand{\ea}{\end{eqnarray}}
\newcommand{\non}{\nonumber \\}

\begin{document}

\title{Dynamical coupled-channel approaches on a momentum lattice}

\author{M.~D\"oring\inst{1}\and J.~Haidenbauer\inst{2,3}\and U.-G.~Mei\ss ner\inst{1,2,3}\and A.~Rusetsky\inst{1}}

\institute{
Helmholtz-Institut f\"ur Strahlen- und Kernphysik and Bethe Center for Theoretical Physics, 
Universit\"at Bonn,\\ Nu\ss allee 14-16, D-53115 Bonn, Germany
\and
Institute for Advanced Simulation and J\"ulich Center for Hadron Physics,
Forschungszentrum J\"ulich, D-52425 J\"ulich, Germany
\and
Institut f\"ur Kernphysik, 
Forschungszentrum J\"ulich, D-52425 J\"ulich, Germany
}

\abstract{  
Dynamical coupled-channel approaches are a widely used tool in hadronic physics
that allow to analyze different reactions and partial waves in a consistent
way. In such approaches the basic interactions are derived within an effective
Lagrangian framework and the resulting pseudo-potentials are then unitarized in
a coupled-channel  scattering equation.  We propose a scheme that allows for 
a solution of the arising integral equation in 
discretized momentum space for periodic as well as anti-periodic boundary
conditions. This permits to study finite size effects as they appear in lattice QCD simulations.  
The new formalism, at this stage with a restriction to $S$-waves,
is applied to coupled-channel  models for the $\sigma(600)$, $f_0(980)$, and
$a_0(980)$ mesons, and also for the $\Lambda(1405)$ baryon. Lattice spectra are
predicted.  
}


\maketitle


\section{Introduction}
\label{Intro}
At low energies chiral perturbation theory has allowed for a comprehensive
understanding of the strong interactions. At higher energies, where the chiral
expansion starts to break down, rich spectra of excited states have been found
in the meson-meson and meson-baryon sectors experimentally,
see e.g. Ref.~\cite{Klempt:2009pi}.
Several analysis tools have been developed to disentangle the partial wave
content and to pin down the resonance spectrum. Among those tools are
dynamical coupled-channel
models~\cite{Krehl:1999km,Gasparyan:2003fp,Doring:2009bi,Doring:2009yv,Doring:2010ap,arXiv:1110.3833,JuliaDiaz:2007kz,Paris:2008ig,Tiator:2010rp},
characterized by interactions driven by hadron exchange,
which are derived from effective Lagrangians. Such coupled-channel models usually fulfill 
two-body unitarity as well as some requirements of three-body unitarity and crossing
symmetry. They are widely used for data analyses nowadays, for example 
by the J\"ulich~\cite{Krehl:1999km,Gasparyan:2003fp,Doring:2009bi,Doring:2009yv,Doring:2010ap,arXiv:1110.3833},
EBAC~\cite{JuliaDiaz:2007kz} and DMT~\cite{Tiator:2010rp} groups in their
investigations of the $\pi N$ system in the second and third resonance region. 

In recent years information about the excited hadron spectrum is becoming
available from the rapidly evolving field of lattice gauge theory.  The masses
of meson and baryon ground states could be determined at the few percent
level~\cite{Durr:2008zz}  and close to physical pion masses.  With regard to
excited hadrons, signals of a rich spectrum could be found
recently~\cite{McNeile:2006nv,Mathur:2006bs,Dudek:2009qf,Bulava:2010yg,Engel:2010my}
in  calculations with pion masses $\geq 300$ MeV or in the quenched
approximation. Of course, those excited states will be able to decay on the
lattice, once the employed pion mass approaches the physical value. 
Then, resonance signals on the lattice manifest themselves in ``avoided level
crossing''; for narrow resonances, this signal is rather clear, but for
broader  resonances the avoided crossing becomes smeared
out~\cite{Bernard:2007cm,Bernard:2008ax}.

In the one-channel- and also two-channel case, and given sufficient precision
for the levels, the phase shift can be extracted unambiguously and without any
model input by using L\"uscher's
formalism~\cite{Luscher:1986pf,Luscher:1990ux,Li:2007ey,Lage:2009zv,Bernard:2010fp,ourmanu}. 
However, the precision of the lattice data is limited. Moreover, at 
higher energies often many more physical channels are open as it is the case, e.g., for
the $\pi N$ interaction in the second and third resonance region.
Thus, model input may be required to extract the physical
properties and the  resonance spectrum, in particular for broad resonances. 

The physically realistic situation of multiple open decay channels
is most conveniently organized in a coupled-channel scheme. Note that this is of 
particular importance in the study of finite size effects as discussed in 
detail, e.g., in Ref.~\cite{ourmanu}. Threshold openings show the same avoided
level crossing as resonances, and it is, thus, necessary to take the physically 
allowed channels consistently into account, at least if they are known to couple
strongly to the system of interest.

In Refs.~\cite{ourmanu,arXiv:1111.0616} a coupled-channel fit to synthetic lattice 
data with given errors has been carried out, stabilizing the extraction of phase shifts
and pole positions by using Unitarized Chiral Perturbation Theory to expand the
potential. The feasibility of this procedure has been shown and the error
propagation from the lattice data to pole positions and phase shifts could be
quantified.

In a similar way, from the point of view of hadron exchange approaches, the
energy levels provided by lattice calculations can be treated as additional data that
enter the combined analysis of hadronic reactions. 
Thus, the motivation of this study is to reformulate the formalism
of the
dynamical coupled-channel approaches in a manner that is suitable for a
prediction of the discrete energy levels on a finite-size lattice.
In particular, we find that the discretization of the multi-channel
Lippman-Schwinger equation in the momentum space allows one to achieve
the goal
with a minimum effort. The paper contains a detailed discussion of such
a discretization procedure.

The proposed framework can be used to produce the volume-dependent
energy spectra for different physical systems.
We are aware that present
day lattice simulations do not provide more than a few data points for a
limited number of volumes. Still, we would like to stress again that, 
as shown in Ref.~\cite{ourmanu}, assuming a
simple and
sufficiently general parameterization for the chiral potentials, it is
possible
to extract the resonance parameters from a fit of the predicted
energy levels to reasonably small data sets.

To demonstrate the discretization of dynamical coupled-channel approaches, in
this study we  utilize hadron-ex\-change potential models that have been
developed to describe the meson-meson interaction in the strangeness $S=0$
sector~\cite{Janssen:1994wn,Krehl:1996rk} and the meson-baryon interaction for
$S=-1$ as, e.g., in Refs.~\cite{MuellerGroeling:1990cw,Haidenbauer:2010ch}. The
meson-baryon sector with
$S=0$~\cite{Krehl:1999km,Gasparyan:2003fp,Doring:2009bi,Doring:2009yv,Doring:2010ap}
is left to future studies.

In the aforementioned models, the interaction is given by $s$-, $t$-, and
$u$-channel exchange of hadrons, which connect all partial waves of the
scattering amplitude. Furthermore, using SU(3) symmetry, one can also
relate different reactions in the same framework, as recently
demonstrated~\cite{Doring:2010ap}. Thus, hadron exchange puts strong
constraints on the coupled-channel amplitudes and provides at the same time a
structured background from $t-$ and $u$-chan\-nel processes.  Together with a
minimal set of $s-$channel processes (``genuine'' resonances) this allows for a
reliable extraction of poles and residues from the analytic continuation of the
amplitude~\cite{Doring:2009yv}.

The considered exchange processes constitute the interaction potential, $V$, 
which is then projected to partial waves that  represent the basis in which the
scattering equation is solved. Omitting channel indices, the scattering
(Lippmann-Schwinger (LS)) equation in the center-of-mass frame reads
\cite{Krehl:1999km,Gasparyan:2003fp,Doring:2009bi,Doring:2009yv,Doring:2010ap,MuellerGroeling:1990cw}
\begin{multline}
T(q'',q')=V(q'',q')\\
+\int\limits_0^\infty dq\,
 q^2\,V(q'',q)\frac{1}{\sqrt{s}-E_a(q)-E_b(q)+i\epsilon}\,T(q,q') \ , 
\label{scattering}
\end{multline}
where $q''\equiv|\vec q''|$ ($q'\equiv |\vec q'|$)
is the modulus of the outgoing (incoming) 
three-momentum that may be on- or
off-shell,  $\sqrt{s}$ is the scattering energy, and $E_a=\sqrt{m_a^2+q^2}$
and  $E_b=\sqrt{m_b^2+q^2}$ are the on-mass shell energies of the intermediate
particles $a$ and $b$.  
Note that Eq.~(\ref{scattering}) is formulated in the partial wave basis, i.e. the
amplitude only depends on the modulus of the incoming, outgoing, and intermediate 
particle momenta. The angular dependence of the full $T$-matrix is, 
for meson-baryon interaction, provided by the Wigner 
$d_{\lambda\lambda'}^J(\vartheta)$-functions
in the partial wave decomposition~\cite{MuellerGroeling:1990cw}, 
where $\vartheta$ is the scattering angle and $\lambda\,(\lambda')$ is the helicity
of the incoming (outgoing) baryon. For meson-meson interaction, 
the angular dependence reduces to the standard Legendre polynomials $P_J(\cos\vartheta)$.
The solution $T$ of the LS equation allows to calculate
the observables.  Note that the pseudo-potential $V$ that appears on-shell
and also half off-shell in  Eq.~(\ref{scattering}) is fixed from the underlying
interaction Lagrangian.  In the following, we
abbreviate the denominator in Eq.~(\ref{scattering}) as $\sqrt{s}-E_{\rm
int}$,  i.e., $E_{\rm int}=E_a(q)+E_b(q)$.

Dynamical coupled-channel models of the $\pi N$ interaction often take into
account  intermediate and final $\pi\pi N$
states~\cite{Krehl:1999km,Gasparyan:2003fp,Doring:2009bi,Doring:2009yv,Doring:2010ap,arXiv:1110.3833,JuliaDiaz:2007kz},
though usually only in an effective way. Still, this complicates the structure
of  Eq.~(\ref{scattering})~\cite{Doring:2009yv}. Thus, for the present study 
we concentrate on physical problems where such three-body channels play only 
a minor role. The two interaction models we consider
here~\cite{Janssen:1994wn,MuellerGroeling:1990cw} do  not contain any effective
two-meson states. Also, note that the spherical symmetry on the lattice is
broken, and partial waves mix;  however, this effect is small for $S$-waves
which mix only with waves of an orbital angular  momentum $L\geq
4$~\cite{Luscher:1990ux}. In this study, we concentrate on the formal
development, while the issue of partial wave mixing and multi-meson
intermediate states is left for further investigations. 


\section{Discretization of the momentum space}
\label{sec:butterbeidiefische}
A standard way of  solving the integral equation~(\ref{scattering}) for $T$ is
to discretize the integral and invert a large matrix in the space of off-shell
momenta. We will show in the following that with a few minor changes this
scheme can be adapted to the discretized momentum space that corresponds 
to finite-size lattices as used in lattice
gauge calculations. Explicitly, the integral term in the one-channel case is
written in terms of the matrices ${\bf V}$, ${\bf G}$, and ${\bf T}$ 
\ba
&&\int\limits_0^\infty dq\, q^2\,
\frac{V(q'',q)\,T(q,q')}{\sqrt{s}-E_{\rm int}}\to \ {\bf V}\,{\bf G}\,{\bf T}, 
\non
&&{\bf V}_{ij}=V(q_i,q_j) \ , \ \ {\bf T}_{ij}=T(q_i,q_j) \ , \non
&&{\bf G}={\rm diag}\,\left(\frac{q_1^2\,w_1}
{\sqrt{s}-E_1},\cdots,\frac{q_n^2\,w_n}{\sqrt{s}-E_n}\right) \ , 
\label{matrices}
\ea
where the $q_i$'s are the sampling points ($i,j=1,\cdots,n$) of the integration
with the associated integration  weights $w_i$, and $E_i$ is $E_{\rm int}$
evaluated at $q_i$.  The multi-channel case is discussed at the end of this
section. The LS in its discretized form,
\ba
{\bf T}={\bf V}+{\bf  V}\,{\bf G}\,{\bf T} \ ,
\label{tbig}
\ea
reads component-wise for $i,j,k=1,\cdots,n$
\ba
{\bf T}_{ij}={\bf V}_{ij}+\sum_{k=1}^n {\bf V}_{ik}\,
\frac{q_k^2\,w_k}{\sqrt{s}-E_k}\,{\bf T}_{kj} \ .
\label{tcompo}
\ea
The solution ${\bf T}$ can now be obtained by matrix inversion, i.e. 
\ba
{\bf T}=(\mathds{1}-{\bf V}\,{\bf G})^{-1}{\bf V} \ .
\label{solutiont}
\ea
For the calculation of phase shifts or observables one needs the solution of
the reaction amplitude ${\bf T}$ at the on-shell momenta.  To obtain the
on-shell result one can augment the matrices above by adding the on-shell 
momentum to the sampling points, i.e., $q_{n+1}\equiv q_{\rm on}$, where
$q_{\rm on}$  fulfills $\sqrt{s} = \sqrt{m_a^2+q^2_{\rm on}} +
\sqrt{m_b^2+q^2_{\rm on}}$.  Then, ${\bf T}_{n+1,n+1}$ is the on-shell
amplitude and ${\bf T}_{i,n+1}$ ($i=1,\cdots, n$) are the half off-shell
amplitudes.  Note that ${\bf G}_{n+1,n+1} = 0$.  Furthermore, the two-body
singularity occurring at the on-shell energy in the LS equation~(\ref{scattering}) 
needs to be treated numerically.  One possibility
consists in rotating the sampling points into the complex
plane~\cite{Aaron:1966zz}.  Another commonly used option is the Haftel-Tabakin
scheme~\cite{HaftelTabakin} in which the singular term
in the evaluation of the integral in Eq. (\ref{matrices}) is subtracted numerically and added
analytically at the on-shell point.

For the discretization on the lattice, none of these two schemes needs to be
employed, as we will see in the following. The lattice allows only for discrete
momenta (periodic boundary conditions) which, for a simple cubic lattice,
results in the substitution
\ba
\int\frac{d^3\vec q}{(2\pi)^3}\,f(|\vec q|)&\to&\frac{1}{L^3}\sum_{\vec n}\,f(|\vec q|) \ ,\non 
\vec q&=&\frac{2\pi}{L}\,\vec n, \quad\vec n\in \mathds{Z}^3 \ .
\label{substitute}
\ea
To apply the discretization to dynamical coupled-channel models, the LS
equation~(\ref{scattering}) is rewritten as a three-di\-men\-sio\-nal integral
and discretized according to Eq.~(\ref{substitute}). The three-dimensional
summation can be further simplified by considering the multiplicities of the
$i$-th neighbors for the simple cubic lattice with periodic boundary
conditions, labeled as (P) in the following. It can be shown that 
these multiplicities are given by
the $\vartheta^{\rm (P)}$-series [$\vartheta^{\rm (P)}(i=0,1,2,\cdots)=1, 6,
12,\cdots$], see Ref.~\cite{OeisNumbers}. The $\vartheta^{\rm (P)}$ series is given by the
coefficients of the Taylor expansion around $x=0$ of the function
$g^{\rm (P)}(x)=[\vartheta_3(0,x)]^3$, 
\ba
g^{\rm (P)}(x)=\sum_{i=0}^\infty \vartheta^{\rm (P)}(i)\, x^i
\label{taylor}
\ea
where $\vartheta_3$ is the elliptic $\vartheta$
function~\cite{OeisNumbers}, 
\ba
\vartheta_3(0,x)=\sum_{k=-\infty}^\infty\,x^{k^2} \ .
\ea  
One obtains the scattering equation in discretized momentum space,
\begin{multline}
T^{\rm (P)}(q'',q')=V(q'',q')\\+\frac{2\pi^2}{L^3}
\sum_{i=0}^\infty\,\vartheta^{\rm (P)}(i)
\frac{V(q'',q_i)\, T^{\rm (P)}(q_i,q')}{\sqrt{s}-E_a(q_i)-E_b(q_i)},\quad 
q_i=\frac{2\pi}{L}\,\sqrt{i} \ .
\label{scattdisc}
\end{multline}
Please note that the sampling points $q_i$ of the summation in
Eq.~(\ref{scattdisc}) are different from those in Eq.~(\ref{tcompo}) ($q_i$ is
the distance to the $i$-th neighbors).  Note also that all quantities in this
equation are real.  Furthermore, $T^{\rm (P)}$ will now have an infinite tower of poles,
corresponding to the discrete spectrum of a system in a finite volume.

Finally, we can formulate convenient replacement rules to obtain Eq.
(\ref{scattdisc}) from existing dynamical coupled-channel models of the form of
Eq.~(\ref{tbig}) as given in
Refs.~\cite{Krehl:1999km,Gasparyan:2003fp,Doring:2009bi,Doring:2009yv,Doring:2010ap,Janssen:1994wn,Krehl:1996rk,MuellerGroeling:1990cw,Haidenbauer:2010ch}. 
Comparing Eq.~(\ref{tcompo}) with Eq.~(\ref{scattdisc}), we immediately obtain
for the integration measure $q_i^2\,w_i$ and for the integration sampling
points at $q_i$
\ba
q_i^2\,w_i&\to&\frac{2\pi^2\,\vartheta^{\rm (P)}(i)}{L^3},
\non
q_i&=&\frac{2\pi}{L}\,\sqrt{i},
\quad 
i=0,1,\cdots \ , 
\label{subst2}
\ea
i.e. the first sampling point is at $q_i=0$ with $\vartheta^{\rm
(P)}(0)=1$.  Note that if $V$ and $T$ do not depend on $q$, as it is the
case,  e.g., in the chiral unitary approaches considered in
\cite{Oller:1997ti,Kaiser:1998fi,Oller:1998hw,Meissner:1999vr,Oset:1997it,Jido:2003cb,GarciaRecio:2003ks},
those quantities factorize from the  integral in Eq. (\ref{matrices}) so that
the integral equation (\ref{scattering}) reduces directly to an algebraic
equation. The discretized amplitude can then can be rewritten, up to order
$e^{-L}$ effects, in terms of the L\"uscher zeta  function ${\cal Z}_{00}$
within a $K$-matrix approach and for coupled channels as developed in
Refs.~\cite{Lage:2009zv,Bernard:2010fp,ourmanu}. The influence of some $e^{-L}$
suppressed effects has been recently discussed in Ref.~\cite{ourmanu}. Also in
the present formalism, with a non-factorizing scattering equation, it would be 
possible to rewrite the discretized LS equation~(\ref{scattdisc}) in terms of ${\cal Z}_{00}$ (up
to effects of order $e^{-L}$).

Hybrid boundary conditions~\cite{Okiharu:2005eg,Suganuma:2005ds,Suganuma:2007uv} 
were introduced to distinguish
scattering states from tightly bound quark-antiquark systems. Similarly, as
proposed in Refs.~\cite{Bernard:2010fp,ourmanu}, twisted boundary conditions
provide the possibility to change thresholds in lattice gauge calculations.
This provides a unique opportunity to study the nature of resonances that lie
close to a
threshold~\cite{Baru:2003qq,Baru:2004xg,Doring:2007rz,Doring:2009uc,Bruns:2010sv} 
like, for example, the $f_0(980)$ with regard to the $\bar KK$ threshold, 
because the twisting moves the threshold while the resonance stays put.  
We realize that it could be quite challenging to implement this idea (including
twisting for the sea quarks) in present-day lattice simulations. 

Twisted 
boundary conditions, e.g. for the strange quark, lead to a change in the
summation of the lattice momenta of the $\bar KK$
channel~\cite{Bernard:2010fp}, 
\ba
\sum_{\vec n}\,f(|\vec q|)\to\sum_{\vec n}\,f(|\vec q+\vec\theta/L|) \ ,
\ea
where $\vec\theta$ is the twisting angle that can be chosen to vary the $\bar KK$
threshold. For the present study of dynamical coupled-channel approaches, it is
possible to formulate a scattering equation similar to Eq.~(\ref{scattdisc}),
but with maximally twisted (i.e. anti-periodic) boundary conditions 
$\vec\theta=(\pi,\pi,\pi)$. For this value of
$\vec\theta$, the summation exhibits again a high symmetry: compared to the
periodic case, the origin of the summation is simply shifted by
$(1/2,1/2,1/2)$, i.e. it is in the center of the cubic unit cell. The
multiplicity of the next neighbors is $8$, there are 24 next-to-next neighbors
and so on. In general, the multiplicity of the $i$-th neighbors is given as 8
times the number of ordered ways of writing $n$ as the sum of 3 triangular
numbers.  
We call this multiplicity $\vartheta^{\rm (A)}$ in the following
and $\vartheta^{\rm (A)}(i=0,1,2,\cdots)=8, 24, 24,\cdots$. It can be shown that this series is
given by the coefficients of the Taylor expansion [cf. Eq.~(\ref{taylor})] around $x=0$ of
the function 
$g^{\rm (A)}(x)=[\vartheta_2(0,\sqrt x)]^3\,x^{-3/8}$, where $\vartheta_2$ is the
elliptic $\vartheta$ function~\cite{OeisNumbers},
\ba
\vartheta_2(0,\sqrt x)=2\,x^{1/8}\sum_{k=0}^\infty\,x^{k(k+1)/2} \ . 
\ea
With the new multiplicities $\vartheta^{\rm (A)}$, and taking into account
that the sampling points $q_i$ of the summation also change,  the scattering
equation with anti-periodic boundary conditions takes the form
\begin{multline}
T^{\rm (A)}(q'',q')=V(q'',q') \\
+\frac{2\pi^2}{L^3}\sum_{i=0}^\infty\,\vartheta^{\rm (A)}(i)
\frac{V(q'',q_i)\, T^{\rm (A)}(q_i,q')}{\sqrt{s}-E_a(q_i)-E_b(q_i)}
\label{scatttwisted}
\end{multline}
with redistributed sampling points $q_i$,
\ba
q_i&=&\frac{2\pi}{L}\,\sqrt{\frac{8i+3}{4}},\quad i=0,1,\cdots \ .
\label{qitwist}
\ea
As before, $q_i$ is the distance to the $i$-th neighbors. Like for the periodic
case, simple substitution rules with respect to the continuum formalism can be
formulated. The integration weights change according to
\ba
q_i^2\,w_i&\to&\frac{2\pi^2\,\vartheta^{\rm (A)}(i)}{L^3}
\label{subtwisted}
\ea
and the distribution of the $q_i$ is given by Eq.~(\ref{qitwist}) for the
simple cubic lattice with anti-periodic boundary conditions.

So far, the discretization has been shown only for the one-channel case. The
multi-channel case leads to no further  complications and the formalism applies
to it without changes. One just has to take into account that the  quantities
appearing in Eq.~(\ref{tcompo}) depend also on the channel, for example  ${\bf
V}_{ij} \rightarrow {\bf V}^{\mu\nu}_{ij}$, where $\mu$ and $\nu$ characterize
the outgoing and incoming channels. 
The sum runs now over both the momenta and the intermediate channels. The matrix ${\bf G}$, cf.
Eq.~(\ref{matrices}), is diagonal in channel space, too. If anti-periodic boundary
conditions are imposed for, e.g., the strange quark  in meson-meson
scattering~\cite{Bernard:2010fp,ourmanu}, the substitution rules of
Eq.~(\ref{subtwisted}) are applied to  the $\bar KK$ channel, while for the
$\pi\pi$ channel the periodic rules of Eq.~(\ref{subst2}) are used.  Note that
in this case one has transition potentials $V^{\mu\nu}(q_i,q_j)$ with periodic
($q_i$, $\mu = \pi\pi$) as well as anti-periodic  ($q_j$, $\nu = \bar KK$) sampling
points together.


\section{Results and Conclusions}
In lattice calculations, the energy levels $E(L)$ can be obtained as a
function of the box size $L$ of the simple cubic lattice, which allows to apply
the L\"uscher formalism~\cite{Luscher:1986pf,Luscher:1990ux} to extract phase
shifts. For the extraction of poles and phase shifts in the multi-channel case, see also
Refs.~\cite{Bernard:2010fp,ourmanu}. Here, we simply predict the levels $E(L)$ 
applying the formalism developed in the previous section. The levels $E(L)$ are
given by the singularities of the scattering equation~(\ref{scattdisc}) or
(\ref{scatttwisted}), i.e. by the zeros of
\ba
\det(\mathds{1}-{\bf V\,G})=0
\ea
where ${\bf V}$ and ${\bf G}$ are now given by applying the substitution rules
of Eq.~(\ref{subst2})  or Eqs.~(\ref{qitwist}) and (\ref{subtwisted}) to  the
quantities of Eq.~(\ref{matrices}), depending on whether periodic or anti-periodic
boundary conditions  are employed. Note that in a practical calculation the
sums in Eqs.~(\ref{scattdisc}) and (\ref{scatttwisted}) need to be cut off at
some $i_{\rm max}$ which is chosen in a way that the regularization schemes
render contributions from $i>i_{\rm max}$ negligibly small (this can always be
achieved). For the potential models studied in the present paper $i_{\rm max}$ had to be
chosen in such a way~\footnote{Note also
that the inverse of $q_{i_{\rm max}}$ corresponds to distances
of around $\simeq 0.07-0.08$~fm. At these distances the form factors 
that provide the regularization render the high-$q$ contributions
so small that the results practically do not change.
Consequently, if the lattice spacing $a$ is taken less or equal
to the above values (which are common values in present-day
lattice simulations), the finite-$a$ artifacts are washed out and the
limit $a\to 0$ can be safely performed, as was done in the present paper 
from the beginning. Still, a comprehensive treatment of systematic 
uncertainties tied to the finite spacing would require to address the actual 
lattice action that generates the levels, but that is
beyond of what can be possibly done in the present framework.}
that $q_{i_{\rm max}} \approx 2.5 - 3$ GeV.

We consider the two dynamical coupled-channel models mentioned in the
Introduction.  Specifically, we utilize the interaction potential developed in 
Refs.~\cite{Janssen:1994wn,Krehl:1996rk} for the $I=0,1,\,S=0$
meson-meson sector for which the coupling between the $\pi\pi$ (or
$\pi\eta$) and $\bar KK$ channels is taken into account.
The $\sigma\equiv f_0(600)$, $f_0(980)$,  and $a_0(980)$ resonances 
all appear dynamically generated, i.e. without the need of corresponding
$s$-channel exchanges. On the other hand, the model does include also a 
genuine resonance in form of an $s$-channel
pole diagram, tentatively called $f_0(1400)$ in Ref.~\cite{Janssen:1994wn}, whose 
bare mass is at 1520~MeV.

In the $I=0,\,S=-1$ meson-baryon sector, the dynamical coupled-channel model 
of Refs.~\cite{MuellerGroeling:1990cw,Haidenbauer:2010ch} is considered. It
includes the channels $\pi \Lambda$, $\pi \Sigma$, and $\bar K N$. Like in the meson-meson sector,
the attraction in coupled channels is so strong that the $\Lambda(1405)$ appears dynamically generated.
In this approach, also the channels  $\bar K \Delta$, $\bar K^* N$, and $\bar K^*
\Delta$ are included.  However, these channels contribute only effectively to
the  direct $\bar K N$ interaction in form of box diagrams
\cite{MuellerGroeling:1990cw}. Below the respective thresholds ($E<1700$~MeV)
of these intermediate states, the discretized sum of Eq.~(\ref{substitute})
converges towards  the integral in the large $L$ limit and, therefore, at this
stage of the exploratory investigation we refrain from an explicit
discretization of the  integrals related to these box diagrams.

\begin{figure}
\includegraphics[width=0.44\textwidth]{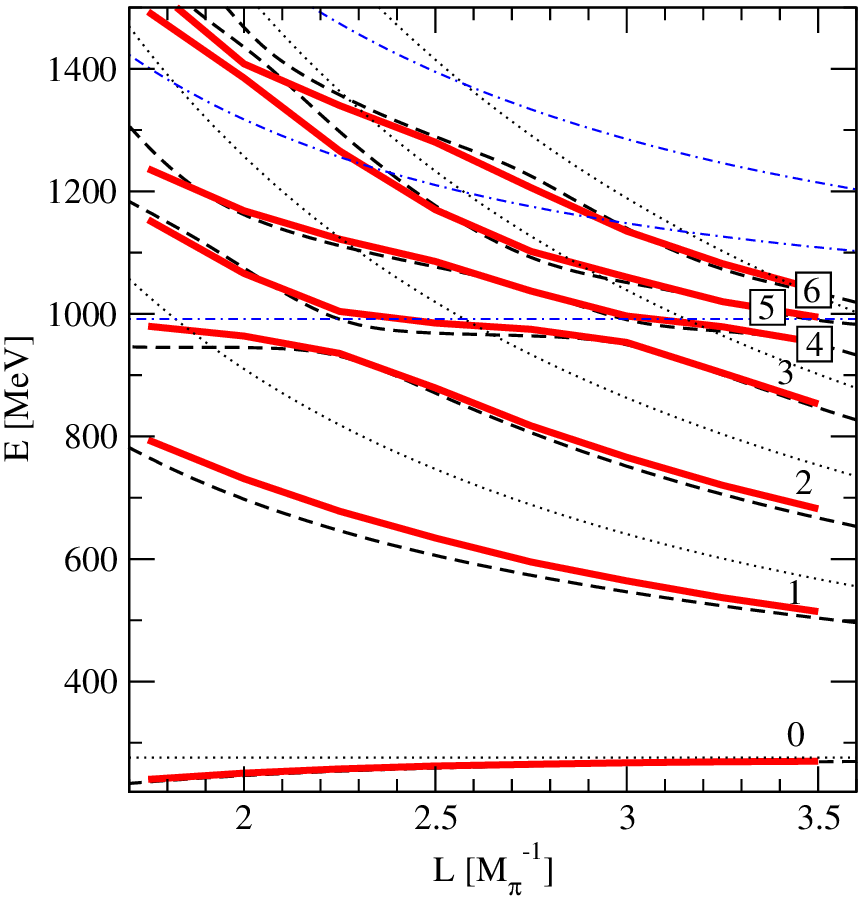}
\caption{Spectrum $E(L)$ for the $I=0$, $J^P=0^+$ meson-meson sector [$\sigma(600)$,
$f_0(980)$] for periodic boundary conditions.  Solid lines: results for the dynamical coupled-channel
approach of  Refs.~\cite{Janssen:1994wn,Krehl:1996rk}.  Dashed lines: For comparison, the results
for the chiral unitary approach of Ref.~\cite{Oller:1997ti},  see also
Refs.~\cite{Bernard:2010fp,ourmanu}. Dotted (dash-dotted) lines: free $\pi\pi\,(\bar KK)$ levels.}
\label{fig:levels_mmI0}
\end{figure}

\begin{figure}
\includegraphics[width=0.44\textwidth]{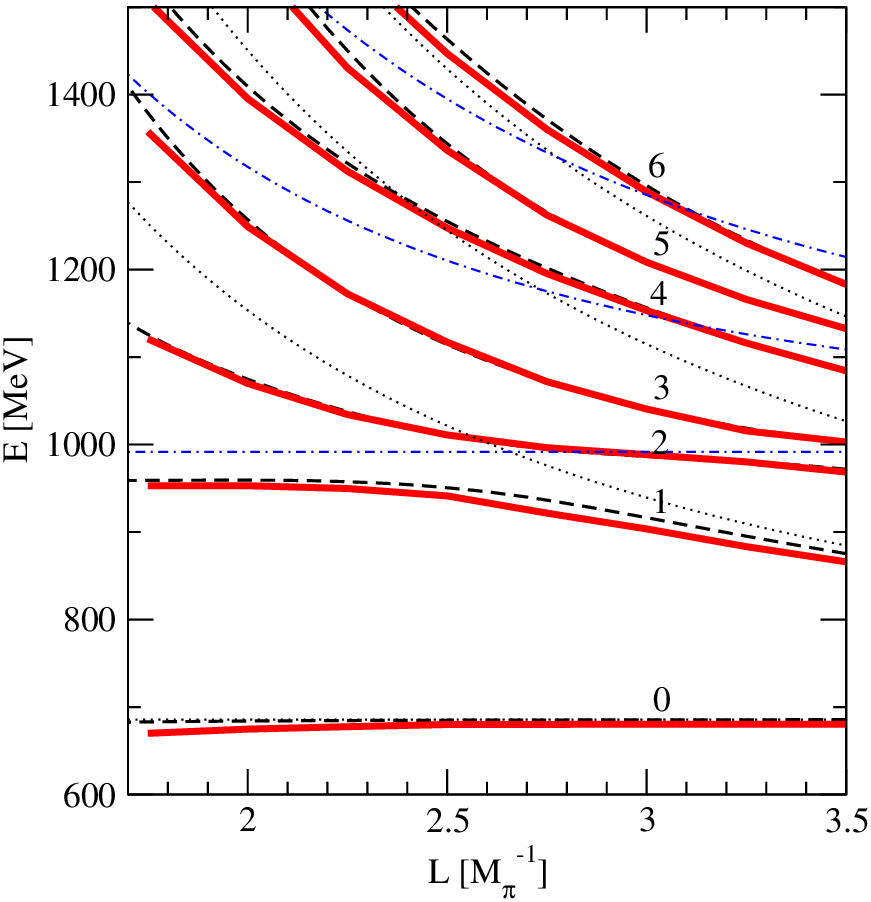}
\caption{Spectrum $E(L)$ for the $I=1$, $J^P=0^+$ meson-meson sector [$a_0(980)$] for periodic boundary conditions. Solid lines: 
results for for the dynamical coupled-channel approach of Refs.~\cite{Janssen:1994wn,Krehl:1996rk}. 
Dashed lines: For comparison, the results
for the chiral unitary approach of Ref.~\cite{Oller:1997ti}.
Dotted (dash-dotted) lines: free $\pi\eta\,(\bar KK)$ levels. 
}
\label{fig:levels_mmI1}
\end{figure}

\begin{figure}
\includegraphics[width=0.44\textwidth]{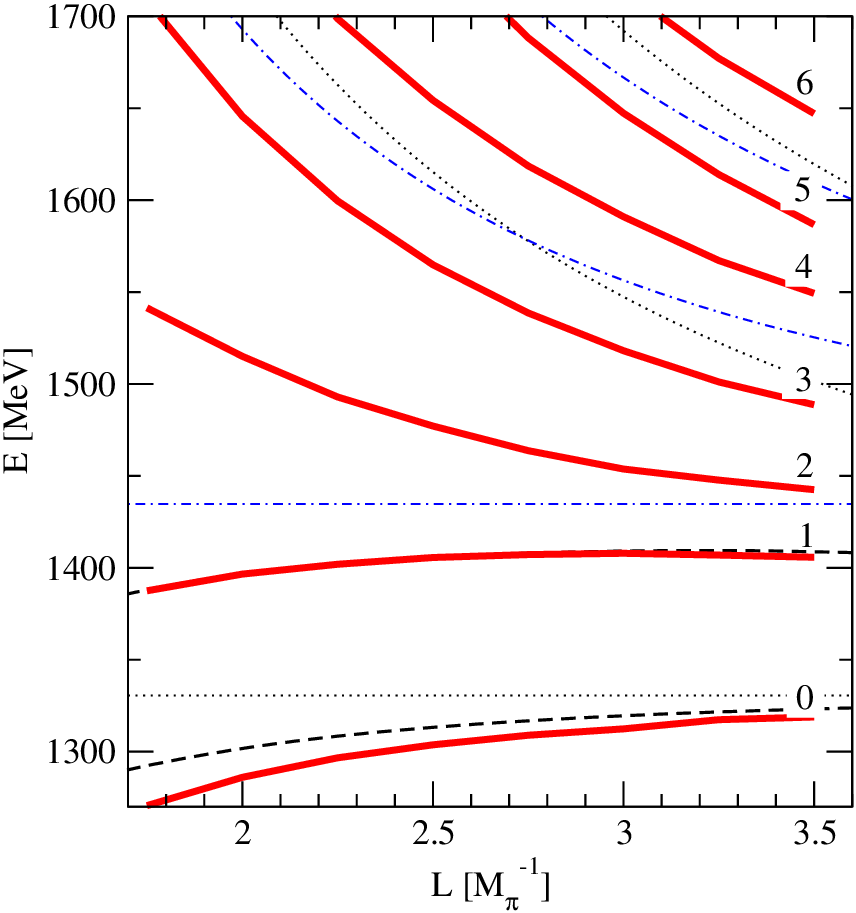}\\
\caption{Spectrum $E(L)$ of the $I=0, \,S=-1$, $J^P={\frac{1}{2}}^-$ meson-baryon sector
[$\Lambda(1405)$] for periodic boundary conditions. Solid lines: results for the $\bar KN$ J\"ulich model of hadron
exchange~\cite{MuellerGroeling:1990cw,Haidenbauer:2010ch}. 
Dashed lines: For comparison, 
the first two levels as obtained from the chiral unitary approach of Ref.~\cite{Oset:1997it}. 
Dotted (dash-dotted) lines: free $\pi\Sigma\,(\bar KN)$ levels.}
\label{fig:lam}
\end{figure}

The calculated levels, using periodic boundary conditions, are shown by the
solid lines in  Figs.~\ref{fig:levels_mmI0}, \ref{fig:levels_mmI1}, and
\ref{fig:lam}, together with the respective free levels of the involved channels.  
In case of the meson-meson interaction the results exhibit a
characteristic avoided level crossing at around  $980$ MeV (more pronounced for
$I=0$ than for $I=1$). In general, avoided level crossing
is a signal for a resonance, but the avoided
crossing is washed out for broader resonances. This applies  for example
to the broad $f_0(600)$ resonance as is obvious from
Fig.~\ref{fig:levels_mmI0}. 
But it is also the case for the genuine $f_0(1400)$ resonance which acquires 
a significant width due to renormalization in the course
of solving the Lippmann-Schwinger equation~(\ref{scattering}).
Its pole is finally at $1346\pm i\,249$~MeV, cf. Table V in Ref.~\cite{Janssen:1994wn},
and thus it is almost as broad as the $f_0(600)$. 
In both cases no direct sign of the presence of these resonances 
is visible in the levels [cf. Fig.~\ref{fig:levels_mmI0}].
Nevertheless, in case of the $f_0(600)$ the pole position can be reconstructed from the levels (with large errors) by expanding 
a general pseudo-potential in powers of the scattering energy, as recently shown in Ref.~\cite{ourmanu}. In the future, 
we plan to address this (inverse) problem of how to reconstruct resonances from a given set of lattice data, in the present framework
of hadron exchange.

Avoided level crossing is also present at particle thresholds. Resonances close to thresholds, like
the $f_0(980)$ that is close to the $\bar KK$ threshold, require, thus, special attention. The avoided crossing,
visible in Fig.~\ref{fig:levels_mmI0}, is therefore not a clear signal for the $f_0(980)$, but necessitates additional
analysis. The inverse problem of reconstructing phase shifts and resonance poles from lattice levels, in such circumstances, has been addressed
recently in Ref.~\cite{ourmanu}, see also Ref.~\cite{Torres:2011pr}.

As mentioned above, the channel space in the $I=0$ sector comprises the $\pi\pi$ and the $\bar KK$ channels. The $4\pi$ channel
has been neglected as there are no phenomenological indications by the PDG~\cite{Nakamura:2010zzi} that require its inclusion at the energies of interest. Still, at energies beyond the $f_0$, the analysis could be refined by including these states, e.g. in terms of quasi-particle channels such as $\sigma\sigma$, $\rho\rho,\cdots$, as carried out, e.g. in Ref.~\cite{Albaladejo:2008es}. In any case, the level structure of 4-body states is considerably more complicated than the one of two-body states considered here, and beyond the scope of this study.

In the meson-baryon sector, cf.
Fig.~\ref{fig:lam}, the level structure below the $\bar KN$ threshold  is 
interesting because of a conjectured two-pole structure of the $\Lambda(1405)$ 
\cite{Jido:2003cb,Oller:2000fj}. Indeed, also the hadron-exchange approach of Ref.~\cite{MuellerGroeling:1990cw}, considered here, predicts two
poles below the  $\bar KN$
threshold~\cite{Haidenbauer:2010ch}. 
Note also the recent work of Ref.~\cite{Menadue:2011pd} in which, for the first time, 
the $\Lambda(1405)$ could be isolated in a lattice QCD calculation near the physical pion mass.

\begin{figure}
\includegraphics[width=0.44\textwidth]{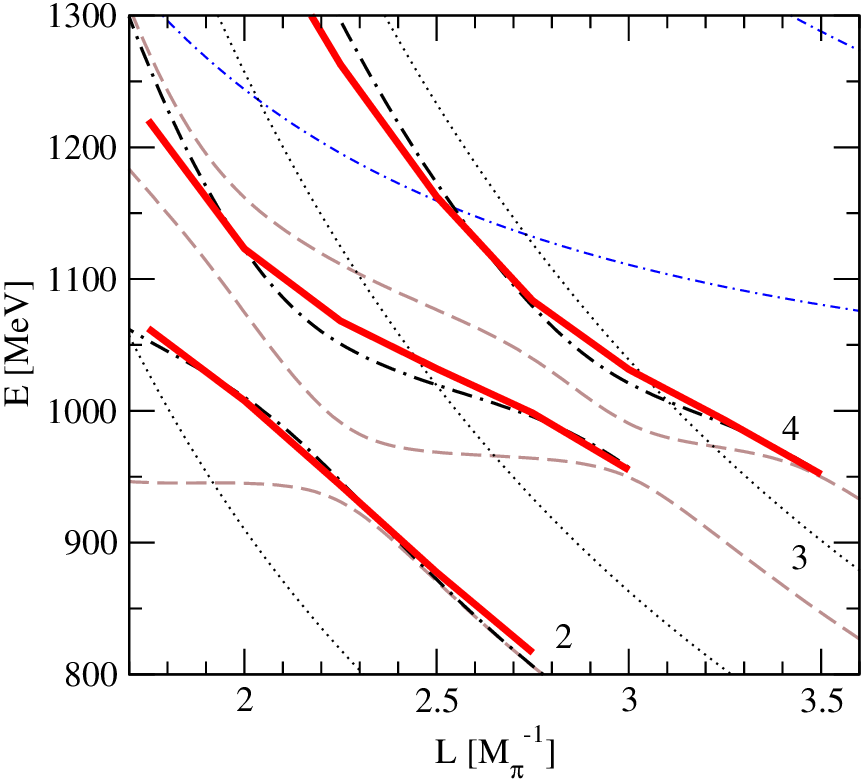}\\
\caption{Same as Fig.~\ref{fig:levels_mmI0} (meson-meson, $I=0$) but with
anti-periodic boundary conditions ($\theta=\pi$) in the $\bar KK$ channel, for the
dynamical coupled-channel model~\cite{Janssen:1994wn,Krehl:1996rk} (solid 
lines). For comparison, the result with anti-periodic boundary conditions for the chiral
unitary approach of Ref.~\cite{Oller:1997ti} is shown (thick dash-dotted lines). Also, the result with periodic boundary conditions of
Fig.~\ref{fig:levels_mmI0} is again shown (dashed lines). Dotted (thin dash-dotted) lines: free $\pi\pi$\,(free anti-periodic $\bar KK)$ levels.}
\label{levels_mmI0_twisted}
\end{figure}

For the $I=S=0$ meson-meson case, we also apply the formalism for anti-periodic
boundary conditions developed in the  previous section. Corresponding result
are presented in Fig.~\ref{levels_mmI0_twisted}.  The bending of the levels at
the $\bar KK$ threshold from avoided level crossing disappears with anti-periodic
boundary conditions as indeed clearly visible in the figure. This
demonstrates again the usefulness of anti-periodic boundary conditions to study
near-threshold resonances like the $f_0(980)$ discussed above~\cite{Bernard:2010fp,ourmanu}.

For comparison, in Figs.~\ref{fig:levels_mmI0}, \ref{fig:levels_mmI1}, 
and \ref{levels_mmI0_twisted} we also show the spectra
calculated in Refs.~\cite{Bernard:2010fp,ourmanu}, using the chiral unitary
approach from Ref.~\cite{Oller:1997ti}.  As
seen in the figures, the levels are
similar to the result based on the hadron-exchange interaction but not
identical, see Fig.~\ref{fig:levels_mmI0}.  Indeed, also the respective phase
shifts in the continuum case show noticeable
differences~\cite{Janssen:1994wn,Oller:1997ti}. The further interpretation of
the  calculated spectra and observed small discrepancies between different
approaches are left for future studies. 

In conclusion, dynamical coupled-channel models can be modified so that the
calculation of lattice spectra becomes possible, in agreement with the
L\"uscher formalism  up to effects of order $e^{-L}$. This framework can be
used to predict lattice spectra, as done here, or to analyze lattice spectra
once lattice data will become available. Dynamical coupled channel approaches,
which respect analyticity, unitarity and other general requirements of the
$S$-matrix, can thus provide the opportunity to analyze and interpret both
experimental data and lattice data within one single approach.

\section*{Acknowledgments}
 This work is supported by the Helmholtz Association by funds provided to
the virtual institute ``Spin and Strong QCD'' (VH-VI-231), by the EU-Research
Infrastructure Integrating Activity ``Study of Strongly Interacting Matter"
(HadronPhysics2, grant n. 227431) under the Seventh Framework Program of the EU,
by the DFG (TR 16 and ``Sachbeihilfe'' GZ: DO 1302/1-2), 
and by COSY FFE under contract 41821485 (COSY 106). 
A.R. acknowledges  support of the Georgia National Science Foundation (Grant
\#GNSF/ST08/4-401).


\end{document}